\documentclass[twocolumn,aps,showpacs,prb,amsmath]{revtex4}
\usepackage{epsfig}
\usepackage{graphicx}
\usepackage{dcolumn}
\usepackage{bm}
\begin{document}
\title{Super-Poissonian noise in a Coulomb blockade metallic
quantum dot structure}
\author{ V. Hung Nguyen and V. Lien Nguyen \footnote{Corresponding
author, E-mail: nvlien@iop.vast.ac.vn}}
 \affiliation{Theoretical Dept., Institute of Physics, VAST
P.O.Box 429 Bo Ho, Hanoi 10000, Vietnam }

\vspace*{1.5cm}
\begin{abstract}
The shot noise of the current through a single electron transistor
(SET), coupled capacitively with an electronic box, is calculated,
using the master equation approach. We show that the noise may be
sub-Poissonian or strongly super-Poissonian, depending mainly on
the box parameters and the gate. The study also supports the idea
that not negative differential conductance, but charge
accumulation in the quantum dot, responds for the super-Poissonian
noise observed.
\end{abstract}


\maketitle

Deviations of the shot noise (SN) from the full (Poissonian) value
in nano-structures have been the subject of a great number of
works, both experimental
\cite{bir,iann,kuzn,haug,gatt,song,safon,alesh,nauen} and
theoretical \cite{hersh,korot,iann,blant,butt,bagre,oriol,hung}.
Mathematically, the measure of these deviations is the Fano factor
$F_n$, defined as the ratio of the actual noise spectral density
to the full SN-value, $2eI$, where $e$ is the elementary charge
and $I$ is the average current. Physically, it is widely accepted
that the Pauli exclusion and the charge interaction are the two
correlations, which cause observed SN-deviations. While the Pauli
exclusion always causes a suppression of SN, the charge
correlation may suppress or enhance the noise, depending on the
conduction regime. The typical non-Poissonian behaviors of SN can
be found in resonant tunneling diodes (RTD), where the noise is
partially suppressed (sub-Poissonian noise) at low bias voltages
(pre-resonance) and becomes very large (super-Poissonian) in the
negative differential conductance (NDC) region
\cite{iann,kuzn,song,blant,alesh}. For Coulomb blockade quantum
dot (QD) structures, a suppression of SN is widely demonstrated,
both experimentally \cite{bir,haug,nauen} and theoretically
\cite{hersh,korot,butt,bagre}. Recently \cite{hung}, we have shown
that in a system of two metallic QDs, coupled in series, the SN is
always sub-Poissonian even in NDC regions. However, in some
particular QD-structure, as that considered in ref. \cite{gatt}, a
positive correlation in the electronic motion through QDs, leading
to a super-Poissonian noise, might be developed. In this paper we
will show that even in a simple structure of a symmetrical single
electron transistor(SET), coupled capacitively to an electronic
box, as that measured in ref. \cite{heij}, the super-Poissonian
noise can be easily realized even in a positive differential
conductance (PDC) region. Furthermore, in consistency with refs.
\cite{song,safon}, our study supports the idea that the charge
accumulation, not NDC, is ultimately responsible for the
super-Poissonian noise observed.

The equivalent circuit diagram of the structure studied is drawn
in Fig.1$(a)$, where the left QD ($D$) forms a SET, while the
right QD ($B$) acts as an electronic box. Two QDs are coupled to
each other by a capacitance $C_m$, but the electron tunneling
between them is forbidden. The current through the SET depends not
only on the bias voltage $V$ and the gate voltage $V_g$, but also
on the charge state in the box. Such a SET-to-box coupling may
produce an NDC as experimentally observed in ref.\cite{heij}.
 \begin{figure}[htp]
 \begin{center}
 \epsfig{width=8.0cm,file=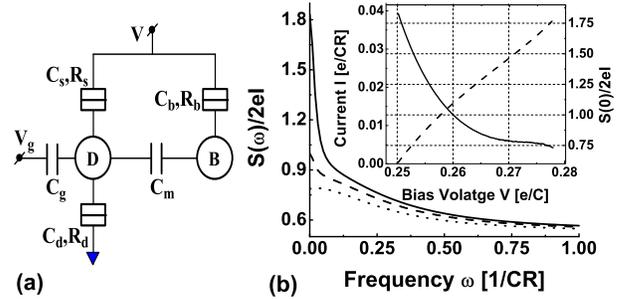,angle=0}
 \caption{$(a)$  Equivalent circuit diagram of the structure under study.
          $(b)$  The normalized noise, $S(\omega )/2eI$, calculated
                 from (8) is plotted as a function of frequency for some values of
                 bias voltages $V $ (from top): 0.25, 0.26, and 0.27. Inset: the current
                 $I$ (7) (dashed line, see the left axis)
                 and the Fano factor (10) (solid line, see the right axis)
                 as a function of bias voltage in the
                 range $V_2 < V < V_3$ (see the text). The
                 structure parameters: $C_m = C_3 = C_2 = C_1 \equiv C$,
                 $R_3 = R_2 = R_1 \equiv R$,
                 without gate.}
 \end{center}
 \end{figure}

Within the framework of the Orthodox theory \cite{aver} the state
$|i>$ of the system under study is entirely determined by the
numbers of excess electrons in two QDs, $n$ in $D$ and $m$ in $B$.
At a given $(n,m)$-state, the free energy of the system can be
written as:
\begin{equation}
\begin{array}{l}
 F = Q_d^2 /2C_d^*  + Q_b^2 /2C_b^*  + Q_d Q_b /C_p^*  \\ 
 \,\,\,\,\,\, - \left( {C_1  + C_3 } \right)V^2 /2\, - C_g V_g^2 /2 - n_q V
\end{array}
\end{equation}
where $C_d^* = \Sigma /C_b ; C_b^* = \Sigma /C_d ; C_p^* = \Sigma
/C_m $ with $\Sigma = C_d C_b - C_m^2 $ , $C_d = C_1 + C_2 + C_m +
C_g , C_b = C_3 + C_m ; Q_d = C_1 V + C_g V_g + n e; Q_b = C_3 V +
me$; and $n_q$ is the number of electrons that have entered the
structure from the top lead (the bottom one is grounded). Any
electron transfer across junctions results in a change in free
energy $F$. In the system of interest there are six possible
sequential electron transfers across three junctions ($1, 2$ and
$3$) upwards (+) or downwards (-). The change in free energy
associated with these transfers can be deduced from eq.(1) as
follows:
\begin{eqnarray}
\begin{array}{l}
 \Delta F_1^ \pm   = e^2 \left( {1 \mp 2n} \right)/2C_d^*  \mp me^2 /C_p^*  \\ 
 \,\,\,\,\,\,\,\,\,\,\,\, \mp eC_g V_g /C_d^*  \mp \left( {C_1 /C_d^*  + C_3 /C_p^*  + 1} \right)eV \\ 
 \Delta F_2^ \pm   = e^2 \left( {1 \pm 2n} \right)/2C_d^*  \pm me^2 /C_p^*  \\ 
 \,\,\,\,\,\,\,\,\,\,\,\, \pm eC_g V_g /C_d^*  \pm \left( {C_1 /C_d^*  + C_3 /C_p^* } \right)eV \\ 
 \Delta F_3^ \pm   = e^2 \left( {1 \mp 2m} \right)/2C_b^*  \mp ne^2 /C_p^*  \\ 
 \,\,\,\,\,\,\,\,\,\,\,\, \mp eC_g V_g /C_p^*  \mp \left( {C_1 /C_p^*  + C_3 /C_b^*  + 1} \right)eV
\end{array}
\end{eqnarray}
At zero temperature the rate of a sequential electron transfer
across any $\nu$-junction ($\nu = 1, 2$ or $3$) is well-known
\cite{aver}:
\begin{equation}
\Gamma_\nu = \theta (- \Delta F_\nu ) \ |\Delta F_\nu |/ (e^2
R_\nu ) ,
\end{equation}
where $\theta$ is the step function, $R_\nu $ is the tunneling
resistance of $\nu$-junction and $\Delta F_\nu$ is the
corresponding change in free energy defined in eq.(2)

Using expressions (2) and (3), in principle, one can solve the
master equation (ME) or perform Monte-Carlo simulation to yield
the current as a function of bias voltage $V$($I-V$
characteristics) and further to calculate the noise. The
Monte-Carlo method is very effective at finite temperature, but it
does not allow us to calculate the noise in the important limit of
low frequency \cite{hung}. In this work we will discuss only the
zero-temperature case, therefore, the ME method should be used.
Denoting $p(i)$ as the probability of the state $|i> \equiv (n_i
,m_i )$ of the system, the ME can be written in the matrix form
\cite{hersh}:
\begin{equation}
d\hat{p}(t) / dt \ = \ \hat{M}\hat{p}(t),
\end{equation}
where $\hat{p}(t)$ is a column matrix of elements $p(i,t)$ and
$\hat{M}$ is an evolution matrix with elements defined as follows:
$M(i,j) = \Gamma^+_2 (j) + \Gamma^-_1 (j)$ (if $n_j = n_i -1$ and
$m_j = m_i$); $\Gamma^-_2 (j) + \Gamma^+_1 (j)$ (if $n_j = n_i +
1$ and $m_j = m_i$); $\Gamma^+_3 (j)$ (if $n_j = n_i$ and $m_j =
m_i + 1$); $\Gamma^-_3 (j)$ (if $n_j = n_i$ and $m_j = m_i - 1$)
and $M(i,i) = -[\Gamma^+_1 (i) + \Gamma^-_1 (i) + \Gamma^+_2 (i) +
\Gamma^-_2 (i) + \Gamma^+_3 (i) + \Gamma^-_3 (i) ]$.

Solving the ME (4) under condition $\sum_i p(i,t) = 1$, we can
further calculate the net current,
\begin{equation}
I(t) = q_1 I_1 (t) + q_2 I_2 (t) + q_3 I_3 (t) ,
\end{equation}
where $I_\nu (t) = e \sum_i [\Gamma^+_\nu (i) - \Gamma^-_\nu (i)]
p(i,t)$ is the statistical average current through $\nu$-junction
($\nu = 1, 2$ or $3$), the factors $q_\nu$ are defined as
\cite{facto}: $q_1 = (C_m C_2 + C_2 C_3 )/\Sigma ; q_2 = (C_m C_1
+ C_m C_3 + C_1 C_3 )/\Sigma$ ; and $q_3 = C_m C_2 /\Sigma $ with
$\Sigma$ given in (1).

Next, the noise spectrum $S(\omega )$ of the current $I$ can be
calculated in the way similar to that developed in
refs.\cite{korot,hung}:
\begin{eqnarray}
\begin{array}{l}
 S\left( \omega  \right) = 2\sum\limits_\nu  {q_\nu ^2 A_\nu  }  + 4e^2 \sum\limits_{\nu \mu } {\sum\limits_{ij} {q_\nu  q_\mu  \left[ {\Gamma _\nu ^ +  \left( i \right) - \Gamma _\nu ^ -  \left( i \right)} \right]} }  \\ 
 \,\,\,\,\, \times B_{ij} \left[ {\Gamma _\mu ^ +  \left( {j|\mu ^ -  } \right)p_{st} \left( {j|\mu ^ -  } \right) - \Gamma _\mu ^ -  \left( {j|\mu ^ +  } \right)p_{st} \left( {j|\mu ^ +  } \right)} \right]
\end{array}
\end{eqnarray}
Here, $A_\nu = e(I_\nu^+ + I_\nu^- )$ with $I_\nu^{\pm} = e\sum_i
p_{st}(i) \Gamma_\nu^{\pm}$; the conditional probability
$p(i\leftarrow j|\tau )$ for having state $|i>$ at the time $t =
\tau > 0$ under the condition that the state was $|j>$ at an
earlier time $t = 0$ obeys the same ME as for the probability
$p(i,t)$; the stationary probability $p_{st}(i)$ is defined as
$p(i\leftarrow j|\tau \rightarrow \infty ) = p_{st}(i)
\delta_{ij}$; $\hat{B} = Re\{ ( i\omega \hat{I} -
\hat{M})^{-1}\}$; $<j|\nu^{\pm} >$ is the state obtained from the
state $|j> = (n_j ,m_j )$ by transferring an electron across the
$\nu$-junction upwards (+)/downwards (-); the tunneling rates
$\Gamma_\nu^{\pm}$ and the factors $q_\nu$ are defined in eqs.(3)
and (5), respectively. Similarly, we can also obtain the noise
expression for currents through junctions, $I_1$ or $I_2$
\cite{hung}.

Thus, using the tunneling rates (3), in principle, we can solve
the ME (4) and further to calculate the current (5) and the noise
(6). In practice, however, this ME can not be exactly solved with
all possible states except some simple cases at low bias voltages.
Let us consider such a simple case, when the SET is symmetrical,
$C_1 = C_2 \equiv C$ and $R_1 = R_2 \equiv R$, and the box
parameters are as follows: $C_3 = C$,  $R_3 = R$, and $C_m$
belongs to the range $(\sqrt{6}-1) C /5 \leq C_m \leq (\sqrt{3} +
1) C$. The gate is neglected. With all these assumptions, in the
way similar to that developed in refs.\cite{hung,jpcm} we can
solve the ME (4) as well as calculate the current (5) and the
noise (6) exactly in some ranges of low bias. Neglecting lengthy,
but elementary, algebraic calculations the final results for the
current can be reviewed as follows: (1) the Coulomb blockade
region has the threshold voltage of $V_0 = (e/2C)(C_m + C)/(5C_m +
3C)$; (2) In the next range of bias voltage, $V_0 \leq V \leq V_1
\equiv (2/2C)(C_m + 2C)/(5C_m + 4C)$, the current has been found
as $I = e \Gamma_2^+ (1) \Gamma_1^+ (0) / [\Gamma_2^+ (1) +
\Gamma_1^+ (0)]$, where two states $|1> \equiv (-1,0)$ and $|0>
\equiv (0,0)$ are written for short; (3) the current is equal to
zero in the range of bias $V_1 \leq V \leq V_2 \equiv (e/2C)(3C_m
+ C)/(5C_m + 3C)$ (second Coulomb blockade gap); and (4) in the
last range of bias, $V_2 \leq V \leq V_3 \equiv (e/2C)(3C_m +
2C)/5C_m + 4C)$, where the ME (3) can be still solved exactly, the
current is given by
\begin{equation}
I = \frac{(a + b) c d h + b c d g}{c d h + b c d + (a + b) d h +
(g + h) c b},
\end{equation}
where we introduce $a = \Gamma_1^+ (0); b = \Gamma_3^+ (0); c =
\Gamma_2^+ (1); d = \Gamma_1^+ (2); g = \Gamma_2^+ (3);$ and $h =
\Gamma_3^- (3)$. Two states $|2> \equiv (0,-1)$ and $|3> \equiv
(1,-1)$ are also written for short.

Now, we can calculate the noise (6) in two ranges of bias, $V_0
\leq V \leq V_1$ and $V_2 \leq V \leq V_3$, where the current is
finite and already known. As an example, we show the noise
expression obtained in the last range of bias, $V_2 \leq V \leq
V_3$:
\begin{equation}
S(\omega ) = 2( q_1^2 A_1 + q_2^2 A_2 + q_3^2 A_3 ) + 4 e^2 D_r B
D_c .
\end{equation}
Here, $A_1 = A_2 = e I$ ($I$ defined in (7)); $A_3 = 2e^2 b c d h
/[ c d h + b c d + (a + b) d h + (g + h) c b ]$; $D_r$ is a
row-matrix of four elements: $q_1 a + q_3 b, q_2 c, q_1 d$ and
$q_2 g - q_3 h$; $D_c$ is a column-matrix of four elements: $q_2
(a + b) c d h / Q, (q_1 a c d h - q_3 b c d h)/Q, (q_2 b c d g +
q_3 b c d h)/Q$ and $q_1 (g + h) b c d / Q$ with $Q = c d h + b c
d + (a + b) d h + (g + h) b c$; and $\hat{B} = Re(i\omega \hat{I}
- \hat{M})^{-1}$ with
\begin{eqnarray}
i\omega \hat{I} - \hat{M} = \left( \begin{array}{cccc}
        i\omega + a + b &   -c        & 0           & 0   \\
       -a               & i\omega + c & 0           & -b  \\
       -b               &      0      & i\omega + d & -g  \\
        0        &      0      &      -d     & i\omega + g + h
           \end{array} \right)
\end{eqnarray}
The expression (8) gives the SN  of the net current $I$ as a
function of frequency $\omega$ and of bias $V$. In Fig.1$(b)$, for
example, we present the normalized noise, $S(\omega )/2eI$, as a
function of frequency, calculated at some bias voltages, for the
structure with parameters given in the figure. Here and below, for
symmetrical SETs of equal capacitances, $C$, and tunneling
resistances, $R$, it is convenient to choose the elementary charge
$e$, the capacitance $C$, and the resistance $R$ as basic units.
The voltage, the current, and the frequency in figures are then
measured in units of $e/C$, $e/C R$, and $(C R )^{-1}$,
respectively. It seems from Fig.1$b$ that the normalized noise
obtained for the net current always decreases as the frequency
increases and within the framework of the model considered  there
exists a large frequency limit: $S(\omega )/2eI \geq 0.5$.

Particularly, in the limit of zero frequency, when all the noises,
for the net current and for currents through partial junctions,
are coincident \cite{hung}, we obtain an explicit expression for
Fano factor:
\begin{eqnarray}
&&F_n  \equiv S\left( 0 \right)/2eI = 1 +  \nonumber \\ 
&&2\{ \frac{{ach\left( {a + h} \right)\left( {d + g + h} \right) + bdg\left( {a + b + c} \right)\left( {g + h} \right)}}{{\left[ {dh\left( {a + b + c} \right) + bc\left( {d + g + h} \right)} \right]\left[ {h\left( {a + b} \right) + bg} \right]}} \\
&&- \frac{{cd\left[ {bc + dh + \left( {a + b + c} \right)\left( {d + g + h} \right)} \right]\left[ {h\left( {a + b} \right) + bg} \right]}}{{\left[ {dh\left( {a + b + c} \right) + bc\left( {d + g + h} \right)} \right]^2 }}\} \nonumber
\end{eqnarray}
where the quantities $a, b, c, d, g$ and $h$ are defined in
eq.(7). Clearly, this expression (10) shows that $F_n$ may be
greater or smaller than 1, depending on relative values of two
terms with opposite signs in the braces. In other words, we have
exactly shown that at least for the simple case under study the SN
may be super-Poissonian or sub-Poissonian, depending on the
structure parameters and bias voltage. Such an interesting noise
behavior can be seen in the inset of Fig.1$(b)$, where, as an
example, we present the current $I$ (7) and the corresponding Fano
factor $F_n$ (10) (valid in the range of bias $V_2 \leq V \leq
V_3$) for the same structure as in the main figure. While the
current monotonously increases (with an PDC), the noise is
super-Poissonian ($F > 1$) at $V \leq 0.26$ and becomes
sub-Poissonian at higher biases.

To extend calculations to higher biases and different varieties of
structure parameters, we solve the ME (4) and calculate the
current (5) and the noise (6) numerically. In Fig.2$(a)$ we
present obtained results of the current $I$ (dashed line) and Fano
factor $F_n$ (solid line) for the structure with the same
parameters as in Fig.1$(b)$ except the SET-to-box capacitance
$C_m$. Apparently, the I-V characteristics obtained is very
similar to that reported in \cite{heij} with a clear second
Coulomb gap. Compared to this experiment, the calculation has been
extended to higher bias voltages, where one more NDC region has
been recognized. Along with such an I-V curve the Fano factor
$F_n$ strongly varies with the bias $V$ and reaches
super-Poissonian peaks, $F_n = 1.31$ and $3.01$, at $V = 0.27$ and
$0.43$, respectively. Note that the lower value of $V$ belongs to
a PDC region, while at the higher one we have an NDC. Statistics
 \begin{figure}[htp]
 \begin{center}
 \epsfig{width=8.7cm,file=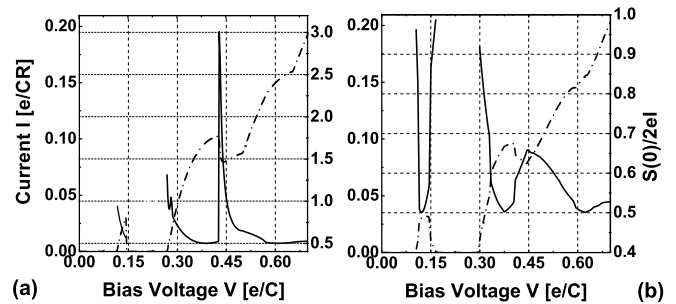,angle=0}
 \caption{Numerical results: the current, calculated from
          (5) (dashed line, see the left axis)
          and the Fano factor, $F_n = S(0)/2eI$,
          calculated from (6) (solid line, see the right axis),
          are plotted against the bias voltage $V$. The
          structure parameters are the same as in Fig.1$b$
          except $C_m$, which is equal to $2 C$ in $(a)$ and $10 C$ in $(b)$}
 \end{center}
 \end{figure}
 \begin{figure}[htp]
 \begin{center}
 \epsfig{width=6.0cm,file=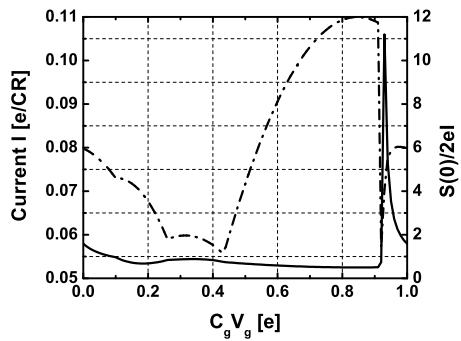,angle=0}
 \caption{The gate effect: the current (dashed line) and
          Fano factor (solid line) are plotted against the gate parameter $C_g
          V_g$ for the same structure as that studied in
          Fig.2$a$ at bias voltage $V = 0.44$.}
 \end{center}
 \end{figure}
of numerical results for structures with different varieties of
parameters show that the noise-versus-bias behavior is very
sensitive to the SET-to-box capacitance $C_m$. A change in $C_m$
can make a super-Poissonian noise sub-Poissonian and inversely.
This can be seen, for example, by comparing two figures,
Fig.2$(a)$ and $(b)$. The structures studied in these figures are
the same except the capacitance $C_m$, which is equal to $2 C$ in
$(a)$ and $10 C$ in $(b)$. While two I-V curves (dashed lines) are
not much different from each other and, particularly, the NDC
regions are still clearly maintained in both figures, in
Fig.2$(b)$ the noise is sub-Poissonian in the whole range of bias
voltages under study. The study demonstrates that by changing only
$C_m$ it is possible to get the noise as large as $F \approx 100$.
Such a giant enhancement of noise has been suggested in a quantum
shuttle at the shutting threshold \cite{novot}.

Results similar to those in Fig.2 have been also obtained when we
change only the box parameter $C_3$ or $R_3 $. Noting again that
the SET is still symmetrical, our study thus demonstrates an
important role of the box in affecting both the I-V
characteristics and the noise behavior of the SET.

All the results presented in Figs.1-2 are for the case without
gate. The gate leads to an additional term in the free energy $F$
(1) and simply makes numerical calculations little lengthier. As
an example, the current (dashed line) and normalized
zero-frequency noise (solid line), calculated at the bias $V =
0.44$, are plotted against the gate parameter $C_g V_g$ in Fig.3
for the same structure as in Fig.2. The Fano factor decreases from
the value of $1.59$ (super-Poissonian) in the case without gate
($C_g V_g = 0)$ to the sub-Poissonian value of $0.5$ at $C_g V_g =
0.85$ and then sharply rises to a large value of $\approx 11.2$.
Note that as the gate parameter $C_g V_g$ varies the changes of
conductance and of noise are not always in accordance with each
other: the noise may be either suppressed or enhanced in NDC
regions. Experimentally, for the structure measured in
\cite{safon}, it was noted that the super-Poissonian peaks can be
observed in only specific ranges of gate voltage.

The fact that a super-Poissonian noise is not necessarily
accompanied by an NDC has been claimed by Song et al. \cite{song}
and by Safonov et al. \cite{safon}. Comparing the I-V curves and
the noises measured in a super-lattice diode and in a RTD, Song et
al. concluded that not NDC, but charge accumulation in the well,
responds for the super-Poissonian noise observed in RTD. Safonov
et al., measuring the noise in resonant tunneling via interacting
localized states, observed a super-Poissonian noise in the range
of bias, where there is no NDC. They have also pointed out that
the effect on noise of the Pauli exclusion principle and the
Coulomb interaction are similar in most mesoscopic systems. For
our structure of study, in solving the ME, we are able to exactly
analyze the charge states of the dot and the box at bias voltages,
where the super-Poissonian peaks are observed. Studies strongly
support the idea \cite{song,safon} that the charge accumulation in
the dot causes the super-Poissonian noise observed.

In conclusion, we have calculated the current and the SN in a SET
capacitively coupled to an electronic box, using the ME approach.
In a particular case we were able to derive exact expressions for
the I-V characteristics as well as the noise as a function of both
frequency and bias voltage. For different varieties of structure
parameters, including the gate, in a large range of bias voltage
the calculation has been performed numerically. The obtained
results show that the noise may be sub-Poissonian or strongly
super-Poissonian, depending mainly on the box parameters and the
gate. The super-Poissonian noise observed in the structure is not
necessarily accompanied by an NDC. The study supports the idea
that not NDC, but charge accumulation in the dot, responds for the
super-Poissonian noise observed. Such an accumulation is
accelerated by charge states in the box.

This work is in part supported by the Natural Science Council of
Vietnam under grant 410901.


\begin{thebibliography}{99}
\bibitem{facto} The expression (4) with factors $g_\nu$ indicated can be easily
                derived from the work done by voltage source
                considering the individual currents $I_\nu (t)$.
\end{thebibliography}
\end{document}